# Genetic-Algorithm-Based Proportional Integral Controller (GAPI) for ROV Steering Control †

Ahsan Tanveer [1,2,*] and Sarvat Mushtaq Ahmad [3,4]

1. Faculty of Mechanical Engineering, Ghulam Ishaq Khan Institute of Engineering Sciences and Technology, Topi 23640, Pakistan
2. Department of Mechanical & Aerospace Engineering, Institute of Avionics and Aeronautics, Air University, Islamabad 44000, Pakistan
3. Control & Instrumentation Engineering Department, King Fahd University of Petroleum and Minerals, Dhahran 31261, Saudi Arabia; sarvat.ahmad@kfupm.edu.sa
4. Interdisciplinary Research Center for Intelligent Manufacturing and Robotics, King Fahd University of Petroleum and Minerals, Dhahran 31261, Saudi Arabia
* Correspondence: ahsantanveer3883@gmail.com or ahsan.tanveer@mail.au.edu.pk
† Presented at the 2nd International Conference on Emerging Trends in Electronic and Telecommunication Engineering, Karachi, Pakistan, 15–16 March 2023.

**Abstract:** This article presents the design and real-time implementation of an optimal controller for precise steering control of a remotely operated underwater vehicle (ROV). A PI controller is investigated to achieve the desired steering performance. The gain parameters of the controller are tuned using the genetic algorithm (GA). The experimental response corresponding to the step waveform for the GA is obtained. A root-locus-tuned PI controller alongside a simulated-annealing-based PI controller (SAPI) is used to benchmark the response characteristics such as overshoot, peak time, and settling time. The experimental findings indicate that GAPI provides considerably better performance than SAPI and the root-locus-tuned controller.

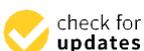

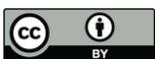

## 1. Introduction

Remotely operated underwater vehicles, or ROVs, are capable of a variety of tasks in both shallow and deep water. They have been effectively used in a variety of contexts, including military operations, commercial endeavours, and research projects, among others. Owing to their electronic subsystem, ROVs can carry out their assigned tasks autonomously while steering effectively in challenging environments. Regardless of the complex, interconnected network of subsystems, it is difficult to operate such vehicles because of environmental disturbances, such as ocean currents.

Researchers throughout the world have used a variety of intelligent control approaches for effective ROV motion control [1,2]. ROV control algorithms are crucial for the efficient and safe operation of these vehicles. Nonetheless, controllers such as PI, PD and PID have exhibited a more simple controlling strategy. Additionally, these approaches have simpler implementation from a computational perspective in the linear domain [3]. The existence of natural disturbances, on the other hand, causes PID controllers to experience arduous computations amid changes in system parameters. There are several articles in the literature that elaborate on the application and design of PID controllers for reliable steering control of the ROV platform [4–6]. In order to ensure accurate regulation of an AUV, this study develops a PI controller that has been optimally tuned using a meta-heuristic optimization technique known as genetic algorithm (GA).

The sections of this article are organised as follows: Section 2 explains the mathematical modelling of an ROV. Section 3 includes the synthesis of the PI controller using GA. The experimental results of the ROV are presented in Section 4. Finally, Section 5 offers a conclusion.

## 2. Dynamic Modelling

ROV's general transfer function, which provides information regarding the yaw angle and its relation with thrust force, is needed in order to develop a PI controller for steering regulation. This is accomplished by the mathematical modelling of ROV [7]. For model development, two separate frames of reference—the earth-fixed and the body-fixed frame—are taken into account. This coordinates of the prototype system are explained using three axes that are mutually perpendicular and start at the centre of gravity of the vehicle, as seen in Figure 1.

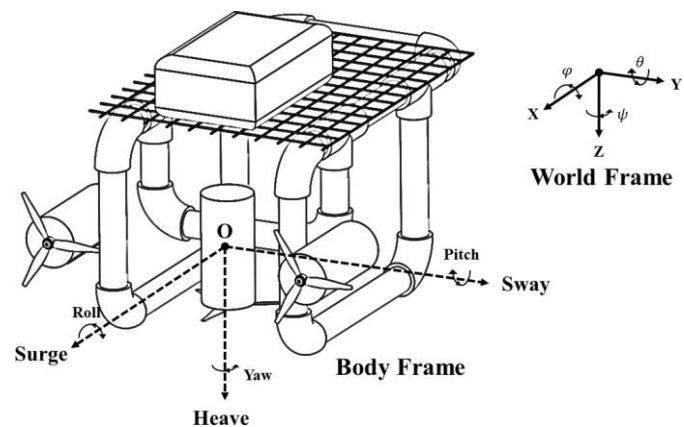

**Figure 1.** Isometric view of the prototype ROV.

In Figure 1, north and east are represented by the *x*-axis and *y*-axis, respectively, whereas depth increases alongside the *z*-axis.

By factoring in the fact that the origin of the body-fixed frame coincides with the centre of gravity, the nonlinear dynamic model of the vehicle is simplified for the pure steering plane:

$$I_z \dot{r} = \Sigma N \quad (1)$$

where *N* represents the net moment in the yaw plane. A thorough modelling exercise has previously been undertaken and may be accessed in [7].

The equation between yaw ($\psi$) and thrust input ($u_t$) in terms of the transfer function is obtained by inserting the values of the vehicle's parameters, thrust and hydrodynamic coefficients, and by applying the Laplace transform [8]:

$$\frac{\psi(s)}{u_t(s)} = \frac{0.01394}{s^2 + 2.08s + 0.4681} \quad (2)$$

Now that a valid transfer function for the ROV has been determined, one may proceed towards the design of the required controller.

## 3. Closed-Loop Control Design

The proportional–integral (PI) is one of the most commonly used algorithms for closed-loop control design. PI controllers have been effectively applied in the design and control of several systems [8,9]. Proportional and integral gain parameters, which are functions of the error between the desired set point and the actual system output, make up a standard PI controller. The characteristic equation is optimised based on the integral time absolute error (ITAE) performance index to determine the exact values of the gain parameters. Finally, GA is employed to optimise the PI controller gains for performance assessment.

*3.1. Controller Tuning Using Meta-Heuristic Optimization*

All dynamical systems frequently employ PI controllers; however, to obtain the desired system response, monotonous tuning of the controller is needed. The repetitive tuning of PI controllers, nonetheless, can be made computationally simpler by using nature-inspired optimization methods. As a result, the ROV's steering controller can be developed with little assistance from humans. In the text that follows, GA is used to fine-tune the PI controller in the steering loop of the ROV. The resultant response is then evaluated against the SAPI controller.

3.1.1. Tuning a PI Controller Using SA

One of the most popular techniques for control problem optimization in dynamical systems is simulated annealing (SA). SA is a stochastic global search optimization algorithm. The metallurgical process of annealing, in which metal is heated to a high temperature quickly and then gently cooled to enhance its strength and make it simpler to work with, serves as an inspiration for the algorithm. The atoms in the material are initially excited at a high temperature, causing them to move around a lot, and then, their excitation is gradually reduced, causing the atoms to fall into a new, low-energy state. The optimal state is the one that exists at this low energy.

Using the SA approach, the gain coefficients were calculated. Following the setup of the optimization problem, the algorithm was performed ten times for the ITAE cost function. This is because, while SA is intended to discover the global minimum of the stated cost function, convergence is ensured for an unlimited number of iterations. As a consequence, the results of the algorithm are presented on a scatter plotter after numerous iterations (as seen in Figure 2), and the controller gains with the lowest cost function value, i.e., $[k_p, k_i] = [296, 81]$, are chosen for simulations and further experimentation.

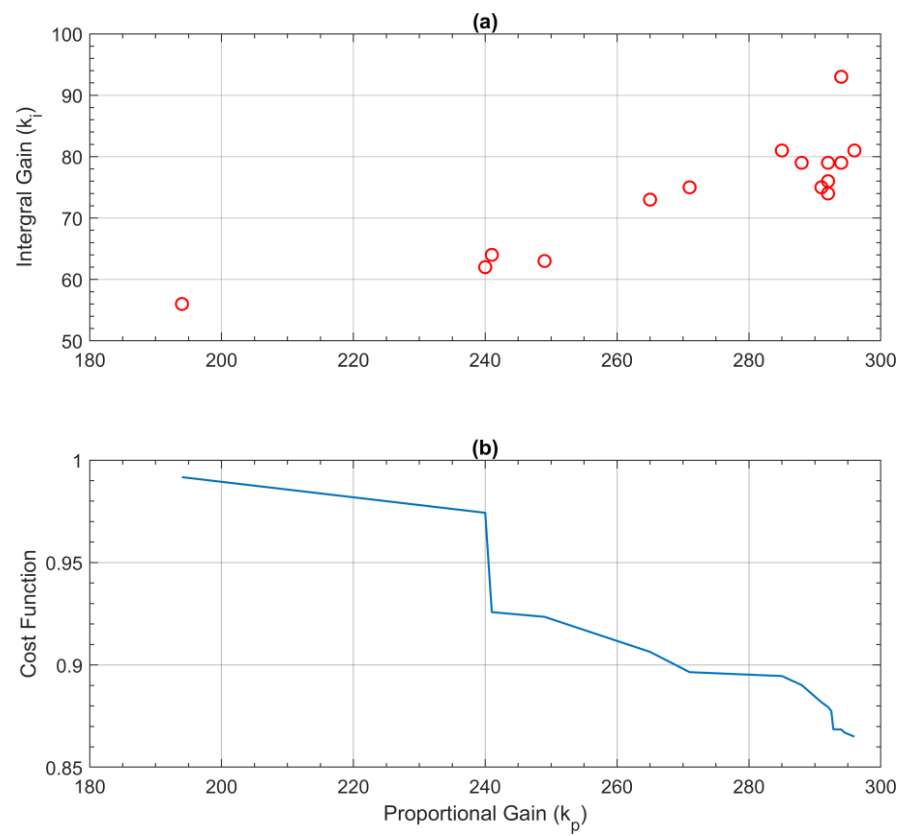

**Figure 2.** SAPI gains for yaw control of the ROVs: (**a**) A scatter plot representing gain parameters obtained after each run of SA optimization. (**b**) A line plot showing value of the cost function associated with particular gain parameter.

### 3.1.2. Tuning a PI Controller Using GA

The genetic algorithm (GA) is an optimization approach based on the phenomenon of natural selection. It is applied to a population of chromosomes in which each chromosome serves as a potential solution with a corresponding fitness value. How ideal is a solution is determined by this value. The starting population is chosen at random, and then, a fitness-based selection method is employed. In order to produce the next generation, recombination is carried out in the subsequent stage. In regeneration, child chromosomes are produced using parent genes. Up until the stopping criterion is realized, the aforementioned iterative processes continues.

Similar to SA, the gain coefficients were obtained after ten iterations of the genetic algorithm for the ITAE cost function. The controller gains with the lowest cost function value, i.e., $[k_p, k_i] = [260, 70]$, were chosen for the simulations and further testing.

## 4. Real-Time Implementation and Results

The output response of the ROV, controlled by tuned PI controllers, was examined for the unit step input in order to compare the performances of GAPI, SAPI, and a conventionally tuned PI controller. The gain parameters for the optimally tuned GAPI, SAPI, and root-locus PI controller were [260, 70], [296, 81], and [230, 90], respectively.

Table 1 lists the transient performance characteristics of all optimally tuned PI controllers.

**Table 1.** Transient performance characteristics of optimally tuned PI controllers in simulation.

| Tuning Strategies | Peak Time (s) | % Overshoot | Settling Time (s) |
|---|---|---|---|
| SA | 1.73 | 21.6 | 4.1 |
| GA | 1.34 | 18 | 4.52 |
| r-locus | 2 | 24 | 3.45 |

Table 1 shows that the performance of the GAPI controller is admirable i.e., having less overshoot and peak time in comparison to SAPI or root-locus-tuned PI. This indicates that the designed GAPI steering controller for the ROV, precisely and optimally fulfills the design requirements.

Once it is established, through simulations, that GAPI is the most optimal controller for the steering problem under discussion, real-time pool experiments were conducted to validate the findings. Figure 3 shows the step response of the tuned controllers.

As evident from Figure 3, GAPI offers a robust reference tracking response. Moreover, the system's response boasts a 1.27 s peak time, a 3.83 s settling time, and a 14% maximum overshoot. Table 2 lists the experimental transient performance characteristics for all tuned PI controllers.

Table 2 makes it obvious that the transient response characteristics for GAPI are far superior to those of the other two controllers. Hence, it can be said that the simulation-predicted response and the experimental response concur and that the GAPI controller results in an optimal performance of the system.

**Table 2.** Transient performance characteristics for all optimally tuned PI controllers in real time.

| Tuning Strategies | Peak Time (s) | % Overshoot | Settling Time (s) |
|---|---|---|---|
| SA | 1.28 | 20 | 4.46 |
| GA | 1.27 | 14 | 3.83 |
| r-locus | 1.28 | 30 | 7.22 |

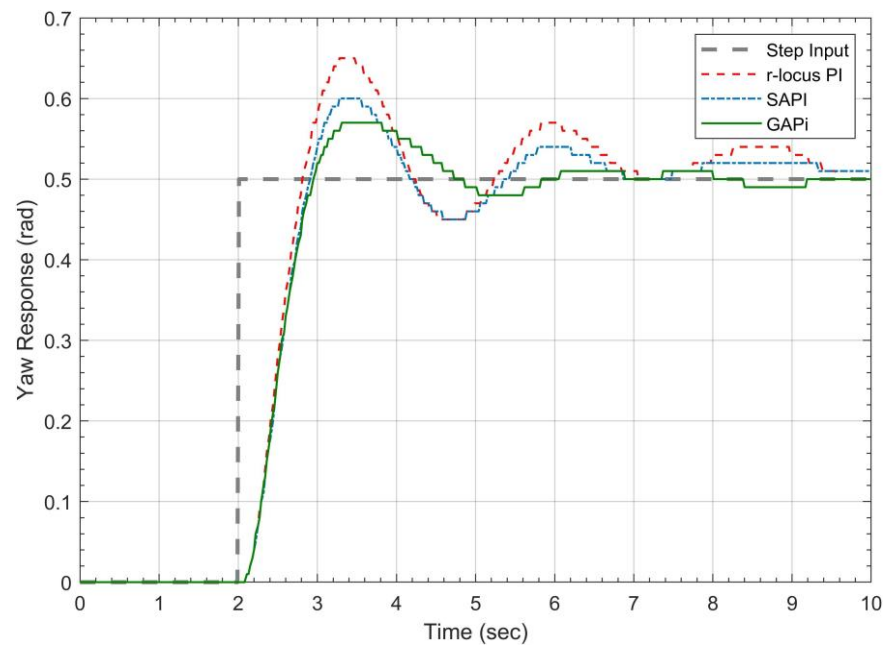

**Figure 3.** Experimental response of the tuned controllers to step input.

## 5. Conclusions

This paper describes the application of a nature-inspired optimization algorithm called genetic algorithm (GA) to optimally tune the gain parameters of a PI controller for effective yaw motion stabilisation of a remotely operated underwater vehicle. The goal was to accomplish robust steering control of the ROV. The suggested PI controller was tuned for an error-based performance metric called ITAE. The step response of the system was benchmarked against the simulated annealing (SA) and root-locus methods. In terms of overshoot, settling time and rise time, the GAPI response was clearly superior to those of SAPI and the r-locus PI. When compared with the r-locus PI, the experimental transient response characteristics showed a 53% improvement in overshoot and a 47% improvement in settling time for GAPI.

In this study, a meta-heuristic optimization of a relatively straightforward PI controller is investigated. The application of the similar optimization techniques to develop more advanced controllers for sophisticated systems will be the focus of future efforts.

**Author Contributions:** Conceptualization, A.T. and S.M.A.; methodology, A.T.; software, A.T.; validation, A.T. and S.M.A.; formal analysis, A.T.; investigation, A.T.; resources, S.M.A.; data curation, A.T.; writing—original draft preparation, A.T.; writing—review and editing, S.M.A.; visualization, A.T.; supervision, S.M.A.; project administration, S.M.A. All authors have read and agreed to the published version of the manuscript.

**Funding:** This research received no external funding.

**Institutional Review Board Statement:** Not applicable.

**Informed Consent Statement:** Not applicable.

**Data Availability Statement:** Data will be made available upon request.

**Acknowledgments:** The authors thank the Faculty of Mechanical Engineering, Ghulam Ishaq Khan Institute of Engineering Sciences and Technology, Topi, Khyber Pakhtunkhwa 23640, Pakistan, for their support for this project.

**Conflicts of Interest:** The authors declare no conflict of interest.